\documentclass[prb,twocolumn,showpacs]{revtex4}
\usepackage{color,soul}
\usepackage{amsmath}
\usepackage{dcolumn}
\usepackage{graphicx}
\usepackage{bm}
\usepackage{amssymb}
\usepackage{subfigure}
\usepackage{diagbox}
\begin{document}

\title{Variational study of the magnetization plateaus of the spin-$\frac{1}{2}$ kagome Heisenberg antiferromagnet and its implication on YCOB}
\author{Rong Cheng and Tao Li}
\affiliation{Department of Physics, Renmin University of China, Beijing 100872, P.R.China}

\begin{abstract}
Numerical simulations find that there are multiple plateaus in the magnetization curve of the spin-$\frac{1}{2}$ Kagome antiferromagnetic Heisenberg model(KAFH) at fractional magnetization $m=1/9,1/3,5/9,7/9$. While it is well known that the $m=1/3,5/9,7/9$ plateau feature a $\sqrt{3}\times\sqrt{3}$ valence bond crystal(VBC) ordering pattern with a David-star-shaped motif, the origin of the narrow plateau at $m=1/9$ remains elusive. Some researchers claim that a subtle translational symmetry breaking pattern with the same $\sqrt{3}\times\sqrt{3}$ periodicity occurs at the $m=1/9$ plateau. On the other hand, it has also been argued that the $m=1/9$ plateau may harbor a novel $Z_{3}$ chiral spin liquid phase. The main challenge to resolve this controversy is to reduce the bias of the method used to describe the magnetization process of such a strongly frustrated quantum magnet, for which intricate competition between different symmetry breaking pattern is generally expected. For this purpose, we have proposed the most general variational ansatz based on the resonating valence bond(RVB) picture that is consistent with the spin symmetry of the system. To optimize such a general RVB ansatz, which can describe all kinds of spatial symmetry breaking pattern on equal footing but at the same time contains a huge number of variational parameters, we have developed a new optimization algorithm, the finite-depth BFGS algorithm. The new algorithm can achieve good balance between numerical efficiency, numerical stability and storage demand and can deal with optimization problem containing millions of parameters at ease. We applied such an advanced optimization algorithm to the general RVB ansatz and mapped out the magnetization curve of the spin-$\frac{1}{2}$ KAFH. We find that a peculiar VBC state with a $3\times3$ periodicity and a windmill-shaped motif has significantly lower energy than the claimed $Z_{3}$ chiral spin liquid state and other proposed VBC states around the $m=1/9$ plateau. We find that there are strong spatial modulation in the local magnetization at the $1/9$ plateau, so strong that even its polarization can be reversed. Our general RVB ansatz also well reproduces all other more conventional magnetization plateaus of the spin-$\frac{1}{2}$ KAFH. We find that the local magnetization is always strongly inhomogeneous below the saturating field for such a strongly frustrated quantum magnet.   
\end{abstract}

\maketitle

\section{Introduction}
The study of the magnetization process of strongly frustrated magnets may reveal more novel state of matters. The most prominent example in this respect is the $1/3$ magnetization plateau in classical antiferromagnetic Heisenberg model defined on the triangular lattice, in which a collinear state with broken translational symmetry is stabilized by the order-by-disorder effect in the intermediate field range\cite{Shender,Kawamura,Chubukov}. In the quantum realm, magnetization plateau may occur for totally different reasons. For example, numerical simulation shows that the spin-$\frac{1}{2}$ antiferromagnetic Heisenberg model defined on the kagome lattice(KAFH) and with nearest-neighboring exchange coupling exhibits well defined magnetization plateaus at fractional magnetization $m=1/3,5/9$ and $m=7/9$\cite{Hida,Richter1,Zhitomirsky,Sakai,Nakano,Hotta1,Richter2,Poiblanc,Richter3}. It is found that the occurrence of these magnetization plateaus are consistent with a generalized Lieb-Schltz-Mattis theorem\cite{Oshikawa}, which states that a featureless gapped phase can only occur when the total magnetization in the unit cell satisfies the requirement $QS(1-m)=integer$, where $Q$ denotes the total number of spins in the unit cell, $S$ is the saturate magnetization of the spin, $m$ is the fractional magnetization per spin at the plateau. Indeed, broken translational symmetry with an enlarged unit cell containing $9$ spins is observed at the $m=1/3,5/9,7/9$ plateau in various numerical simulations of the spin-$\frac{1}{2}$ KAFH. In other words, the occurrence of the magnetization plateau at these fractional magnetization can be understood as a few body effect involving $9$ spins.  More specifically, it is found form numerical simulations that the $m=1/3,5/9,7/9$ plateau all feature a $\sqrt{3}\times\sqrt{3}$ valence bond crystal(VBC) pattern with a David-star-shaped motif\cite{Hotta1}.

From this perspective, it is rather unexpected that there exists an extra narrow plateau at $m=1/9$ in the field range of $B\in[0.3,0.4]$\cite{Hotta1,Poiblanc}. Controversial suggestions on the origin of such an extra plateau has been proposed in the literature. If translational symmetry is still intact, as is assumed in some studies, a gapped spin liquid phase with topological degeneracy is very likely to be realized at such a plateau. The topological degeneracy is here necessary to reconcile the featureless nature of the spin liquid phase with the generalized Lieb-Schultz-Mattis argument at such a fractional magnetization. Early on, a $Z_{3}$ chiral spin liquid phase has been proposed based on the result of grand canonical DMRG simulation of the spin-$\frac{1}{2}$ KAFH\cite{Hotta1,Hotta2,Hotta3}. The $Z_{3}$ chiral spin liquid phase is a bosonic analogy of the fractional quantum Hall state on the lattice and support chiral edge state. Such a proposition is recently supported by a variational study of the spin-$\frac{1}{2}$ KAFH, in which the $Z_{3}$ chiral spin liquid phase is claimed to be stable in a rather broad field range of $B\in[0.35,0.65]$\cite{JXLi}. On the other hand, it has also been proposed that the translational symmetry may have been broken in a subtle way and some kind of valence bond crystal(VBC) phase is realized at the 1/9 plateau\cite{Poiblanc,Fang,Morita}. 

The interest in the $1/9$ magnetization plateau is refueled by the recent experimental discovery made on a new kagome spin liquid candidate material YCu$_{3}$(OH)$_{6}$Br$_{2}$[Br$_{1-x}$(OH)$_{x}$](YCOB)\cite{XHChen,YLi1,SLi2,YLi2}. Unlike its cousin Herbertsmithite ZnCu$_{3}$(OH)$_{6}$Cl$_{2}$\cite{Mendels,Helton,Han} or Zn-Barlowite ZnCu$_{3}$(OH)$_{6}$FrBr\cite{Shi}, which suffer from the subtle issue of Cu-Zn site mixing, the kagome plane in YCOB is thought to be more ideal, since the Y$^{3+}$ ion in YCOB has a radius very different from that of Zn$^{2+}$ in Herbertsmithite. It is found that YCOB remains paramagnetic down to 50mK, even though the exchange coupling strength is about 80K. This indicates that YCOB may constitute a good candidate of quantum spin liquid.  It is important to note that there remains the controversy on the disorder level of the exchange constant in the kagome plane of YCOB, since random replacement of Br by OH ion above and below the Cu hexagons of the kagome plane is unavoidable\cite{YLi1}. 

With these caveats in mind, several groups have measured the magnetization curve in intense magnetic field\cite{Choi,Matsuda1,PALee1,Matsuda2,PALee2}. Thanks to the relatively small exchange constant of YCOB in the kaogme family, it is possible to explore a significantly larger part of the magnetization curve of the spin-$\frac{1}{2}$ KAFH with YCOB. In particular, with a magnetic field of 70T, it is possible to cover the $1/9$ plateau which occurs around $H\sim 0.35J$ and part of the $1/3$ plateau which starts around $H\sim0.8J$. Indeed, both plateaus have been partially confirmed by the observation of significant drop in the differential magnetic susceptibility $dM/dH$ around the field range expected for these plateaus. In addition, it is reported that there is oscillation in the second derivative of the measured magnetic torque $d^{2}\tau/dH^{2}$ in the field range between the $1/9$ plateau and the $1/3$ plateau\cite{PALee1}. Unlike the conventional quantum oscillation in a fermi liquid metal, the oscillation in YCOB is found to be uniform in $B$ rather than uniform in $1/B$. This observation is interpreted as the evidence in support of the emergent fermionic spin excitation and gauge field\cite{PALee1,PALee2}.

While the $1/3$ plateau is more clearly seen experimentally and better understood theoretically, the $1/9$ plateau is more elusive both experimentally and theoretically. This can be partially attributed to the nearly local nature of the $1/3$ plateau state, which makes it more robust against the potential disorder effect. On the other hand, such a local picture is now absent for the $1/9$ plateau state. In DMRG simulation using grand canonical trick, it is found that the $1/9$ plateau state harbors intriguing quasi-degenerate translational symmetry breaking patterns\cite{Hotta1}. This implies that the $1/9$ plateau state is far from being deep inside a gapped and featureless spin liquid phase. At the same time, simulations using the tensor network algorithm find that various translational symmetry breaking phase may be realized at the $1/9$ plateau\cite{Poiblanc,Fang}. The real space pattern found in these tensor network simulations is quite complicated and may depend sensitively on the details in the realization of the tensor network algorithm. This is in some sense in accordance with the findings made from DMRG simulation, which claims that the stability of the quasi-degenerate translational symmetry breaking patterns depend sensitively on the size and the shape of the cluster used in the simulation\cite{Hotta1}. 

With these results in mind, it is then quite suspicious that a fully gapped $Z_{3}$ chiral spin liquid phase can be stabilized at the $1/9$ plateau in so broad a field range of $H\in[0.35,0.63]$, as is claimed in the recent variational study of the spin-$\frac{1}{2}$ KAFH\cite{JXLi}. In addition, the same variational study also predicts that the $1/9$ plateau is  sandwiched by the Dirac spin liquid phase at lower field and the $1/3$ plateau state at higher field with almost no transition region between the plateaus. The resultant magnetization curve is very different from those obtained by other numerical methods.

Reducing variational bias is crucial to reach reliable result in any variational study. This is especially important when we are dealing with the magnetization process of a strongly frustrated quantum magnets, for which intricate competition between different symmetry breaking patterns is generally expected. In the variational study conducted in Ref.\onlinecite{JXLi}, a $U(1)$ RVB ansatz with only nearest-neighboring(NN) RVB parameters and a unit cell containing at most 9 sites is assumed. Although such an ansatz already contains dozens of parameters, it may still be too restrictive to describe the real magnetization process of the spin-$\frac{1}{2}$ KAFH. In fact, if the $Z_{3}$ chiral spin liquid phase is indeed stable in so broad a field range, there would be no reason why the more sophisticated DMRG and the tensor network simulation method fail to identify it in the corresponding field range.

Recently, we have proposed an improved algorithm to perform unrestricted variational optimization within the RVB scheme\cite{Yang1}. The key aspect of the new algorithm is to approximate the Hessian matrix with a finite-depth BFGS iteration. In this paper, we develop further this algorithm to make possible large scale unrestricted optimization involving as many as millions of variational parameters. To describe the magnetization process of the spin-$\frac{1}{2}$ KAFH, we have proposed the most general RVB ansatz consistent with the spin symmetry of the system. Through large scale optimization on such a general RVB ansatz, which contains the NN-$U(1)$ ansatz used in Ref.\onlinecite{JXLi} as a subset, we find that the variational energy of the system is significantly lower than that obtained from both previous variational study and tensor network simulations. The energy difference is especially large in the $1/9$ plateau regime in which the $Z_{3}$ chiral spin liquid phase is claimed. We find that the $1/9$ plateau state features a windmill-shaped VBC pattern with a $3\times3$ enlarged unit cell. Such a VBC pattern is totally different from the $\sqrt{3}\times\sqrt{3}$ VBC pattern anticipated before from tensor network simulation\cite{Fang} and ED analysis\cite{Morita}. 

Our general RVB ansatz can also well reproduce the other three magnetization plateaus of the spin-$\frac{1}{2}$ KAFH. Unlike the $1/9$ plateau state, we find that the $1/3$, $5/9$ and $7/9$ plateau state all feature the well known $\sqrt{3}\times\sqrt{3}$ VBC pattern with a David-star-shaped motif. In addition, we find that the magnetization of the system is always strongly inhomogeneous at a general magnetization. More specifically, we find that the relative inhomogeneity in the local magnetization decreases almost monotonically with increasing field, until it vanishes at the saturation field. In the weak field regime, we find that the relative inhomogeneity diverges as $1/\sqrt{m}$, where $m$ is the average magnetization. 
 
The paper is organized as follows. In the next section, we introduce the model studied in this work and the general RVB ansatz used to describe its magnetization process. In Sec.III, we introduce the improved BFGS algorithm to perform the large scale unrestricted optimization on such a general RVB ansatz. In Sec.IV, we present the numerical results of our variational calculation. The last section of the paper is devoted to the conclusion of our results and a discussion on the relevance of our results to the experimental observations made on the kagome spin liquid candidate material YCOB.

\section{The spin-$\frac{1}{2}$ KAFH model and the variational description of its magnetization process}
The Hamiltonian of the spin-$\frac{1}{2}$ KAFH model in the presence of an external magnetic field is given by 
\begin{equation}
H_{J}=J\sum_{<i,j>}\mathbf{S}_{i}\cdot\mathbf{S}_{j}-B\sum_{i}\mathbf{S}_{i}^{z}
\end{equation}
Here the sum is over nearest-neighboring bonds of the kagome lattice and we assume that the external magnetic field is directed in the $z-$direction. The model thus preserves the spin rotational symmetry along the $z-$direction and the total spin of the system in the $z-$direction is a conserved quantity. In the following, we will set $J=1$ as the unit of energy.

To describe the ground state of the system in the RVB scheme, we represent the spin operator as 
\begin{equation}
\mathbf{S}_{i}=\frac{1}{2}\sum_{\alpha,\beta=\uparrow,\downarrow}f^{\dagger}_{i,\alpha}\bm{\sigma}_{\alpha,\beta}f_{i,\beta}
\end{equation}
in which $f_{i,\alpha}$ is a fermionic slave particle operator defined on site $i$. $\bm{\sigma}$ is the usual spin Pauli matrix. This representation becomes exact when the following constraint
\begin{equation} 
\sum_{\alpha=\uparrow,\downarrow}f^{\dagger}_{i,\alpha}f_{i,\alpha}=1
\end{equation}
is enforced. The fermionic RVB state that we will adopt to describe the ground state of the model is generated from Gutzwiller projection of the mean field ground state of the following BCS-type Hamiltonian
\begin{eqnarray}
H_{\mathrm{MF}}&=&\sum_{i,j,\alpha=\uparrow,\downarrow}(\chi_{i,j}^{\alpha}f^{\dagger}_{i,\alpha}f_{j,\alpha}+h.c.)\nonumber\\
&+&\sum_{i,j}(\Delta_{i,j}f^{\dagger}_{i,\uparrow}f^{\dagger}_{j,\downarrow}+h.c.)
\end{eqnarray}
here the hopping parameter $\chi_{i,j}^{\alpha}$ and the pairing parameter $\Delta_{i,j}$ are the variational parameters in the RVB ansatz and both are assumed to be complex. We also assume that 
\begin{eqnarray}
\chi_{i,j}^{\uparrow}&\neq&\chi_{i,j}^{\downarrow}\nonumber\\
\Delta_{i,j}&\neq&\Delta_{j,i}
\end{eqnarray}
so that only the spin rotational symmetry along the $z$-direction is preserved, as is the case when the external field is applied. At the same time, no assumption on the range of the hopping parameter $\chi_{i,j}^{\alpha}$ and the pairing parameter $\Delta_{i,j}^{\alpha}$ is made and both $\chi_{i,j}^{\alpha}$ and $\Delta_{i,j}^{\alpha}$ can be arbitrarily long-ranged. Thus $H_{\mathrm{MF}}$ actually represents the most general RVB mean field ansatz compatible with the spin symmetry of the model. We note in addition that the ansatz already contains on-site term of the form 
\begin{equation}
\sum_{i,\alpha=\uparrow,\downarrow}\mu_{i}^{\alpha}f^{\dagger}_{i,\alpha}f_{i,\alpha}+\sum_{i}(\Delta_{i}f^{\dagger}_{i,\uparrow}f^{\dagger}_{i,\downarrow}+h.c.)
\end{equation}
The mean field ansatz as given by Eq.4 involves a huge number of variational parameters. On a finite cluster with $N$ site, the total number of parameters is of the order of $4N^{2}$. Such large number of variational parameter are introduced to reduce the bias in our variational study. As we will see later, this is crucial when we are studying the magnetization process of strongly frustrated quantum magnet such as the spin-$\frac{1}{2}$ KAFH.

Denoting the ground state of $H_{\mathrm{MF}}$ as $|\mathrm{MF}\rangle$, the RVB state describing the ground state of the spin-$\frac{1}{2}$ KAFH model can be written as
\begin{equation}
|f-\mathrm{RVB}\rangle=P_{G}|\mathrm{MF}\rangle
\end{equation}
here $P_{G}$ is the Gutzwiller projection operator that removes the doubly occupied configuration from the mean field ground state. To solve the ground state of $H_{\mathrm{MF}}$, we perform the following particle-hole transformation on the down-spin fermion operator
\begin{equation}
f^{\dagger}_{i,\downarrow}\rightarrow \tilde{f}_{i,\downarrow}
\end{equation}
and rewrite the mean field ansatz as the following matrix form
\begin{equation}
H_{\mathrm{MF}}=\bm{\psi}^{\dagger}\mathbf{M}\bm{\psi}
\end{equation}
in which 
\begin{equation}
\bm{\psi}^{\dagger}=(f^{\dagger}_{1,\uparrow},...,f^{\dagger}_{N,\uparrow},\tilde{f}^{\dagger}_{1,\downarrow},....,\tilde{f}^{\dagger}_{N,\downarrow})
\end{equation}
The matrix $\mathbf{M}$ is given by
\begin{equation}
\mathbf{M}=\left(\begin{array}{cc}\chi^{\uparrow}_{i,j} & \Delta_{i,j} \\ \Delta^{*}_{i,j} & -\chi^{\downarrow}_{j,i} \end{array}\right)
\end{equation}
$H_{\mathrm{MF}}$ can then be diagonalized by the following unitary transformation
\begin{equation}
\bm{\psi}=\mathbf{U}\bm{\gamma}
\end{equation} 
here 
\begin{equation}
\bm{\gamma}^{\dagger}=(\gamma^{\dagger}_{1},...,\gamma^{\dagger}_{N},\gamma^{\dagger}_{N+1},....,\gamma^{\dagger}_{2N})
\end{equation}
We note that each $\gamma^{\dagger}_{i}$ carries spin quantum number $+1/2$ in $z-$direction. The columns of the unitary matrix $\mathbf{U}$ are composed of the eigenvectors of the matrix $\mathbf{M}$ in ascending order in its eigenvalue. The mean field ground state then reads
\begin{equation}
|\mathrm{MF}\rangle=\prod_{i=1}^{N+2M}\gamma^{\dagger}_{i}|\tilde{0}\rangle
\end{equation}
in which 
\begin{equation}
M=\sum_{i}\mathbf{S}^{z}_{i}
\end{equation}
in the total magnetization, $|\tilde{0}\rangle$ is the vacuum state of $f_{i,\uparrow}$ and $\tilde{f}_{i,\downarrow}=f^{\dagger}_{i,\downarrow}$, namely the state satisfying 
\begin{equation}
f_{i,\uparrow}|\tilde{0}\rangle=\tilde{f}_{i,\downarrow}|\tilde{0}\rangle=f^{\dagger}_{i,\downarrow}|\tilde{0}\rangle=0
\end{equation}
$|\tilde{0}\rangle$ can also be expressed in terms of the vacuum state of $f_{i,\uparrow}$ and $f_{i,\downarrow}$ as
\begin{equation}
|\tilde{0}\rangle=\prod_{i=1}^{N}f^{\dagger}_{i,\downarrow}|0\rangle
\end{equation}
here $|0\rangle$ satisfies
\begin{equation}
f_{i,\uparrow}|0\rangle=f_{i,\downarrow}|0\rangle=0
\end{equation}
In this work we will always use $|\tilde{0}\rangle$ as our reference state.

To perform variational Monte Carlo simulation on the RVB state $|f-\mathrm{RVB}\rangle$, we expand it in the Ising basis as follows
\begin{equation}
|f-\mathrm{RVB}\rangle=\sum_{\{i_{1},...,i_{N^{\uparrow}_{e}}\}}\Psi(i_{1},...,i_{N^{\uparrow}_{e}})\prod^{N^{\uparrow}_{e}}_{k=1}f^{\dagger}_{i_{k},\uparrow}\tilde{f}^{\dagger}_{i_{k},\downarrow}|\tilde{0}\rangle
\end{equation}
here $N^{\uparrow}_{e}$ denotes the number of up spin site in the system and it satisfies 
\begin{equation}
2N^{\uparrow}_{e}=N+2M
\end{equation}
We note that the Ising basis in Eq.(19) can also be written in terms of the spin raising operator as follows
\begin{equation}
\prod^{N^{\uparrow}_{e}}_{k=1}f^{\dagger}_{i_{k},\uparrow}\tilde{f}^{\dagger}_{i_{k},\downarrow}|\tilde{0}\rangle=\prod^{N^{\uparrow}_{e}}_{k=1}f^{\dagger}_{i_{k},\uparrow}f_{i_{k},\downarrow}|\tilde{0}\rangle=\prod^{N^{\uparrow}_{e}}_{k=1}\mathbf{S}^{+}_{i_{k}}|\tilde{0}\rangle
\end{equation}
The wave function of the RVB state in such an Ising basis is given by
\begin{equation}
\Psi(i_{1},...,i_{N^{\uparrow}_{e}})=\mathrm{Det}[\bm{\Phi}]
\end{equation}
in which $\bm{\Phi}$ is a $2N^{\uparrow}_{e}\times2N^{\uparrow}_{e}$ matrix of the following form
\begin{equation}
\bm{\Phi}=\left(\begin{array}{cccc}U_{i_{1},1} & . & . & u_{i_{1},2N^{\uparrow}_{e}} \\. & . & . & . \\U_{i_{N^{\uparrow}_{e}},1} & . & . & U_{i_{N^{\uparrow}_{e}},2N^{\uparrow}_{e}} \\U_{i_{1}+N,1} & . & . & U_{i_{1}+N,2N^{\uparrow}_{e}} \\. & . & . & .\\U_{i_{N^{\uparrow}_{e}}+N,1} & . & . & U_{i_{N^{\uparrow}_{e}}+N,2N^{\uparrow}_{e}}\end{array}\right)
\end{equation}
in which $U_{i,j}$ is the matrix element of the matrix $\mathbf{U}$. 

With the wave function $\Psi$ in hand, we can now compute the variational energy of the RVB state, which is given by 
\begin{equation}
E=\langle H \rangle_{\Psi}=\frac{\langle\Psi| H |\Psi\rangle}{\langle \Psi |\Psi \rangle}=\frac{\sum_{R}|\Psi(R)|^2 E_{loc}(R)}{\sum_{R}|\Psi(R)|^2}
\end{equation}
here we use $|R\rangle$ to denote a general Ising basis vector. $E_{loc}(R)$ is the local energy of this basis and is defined by
\begin{equation}
E_{loc}(R)=\sum_{R'}\langle R |H| R' \rangle \frac{\Psi(R')}{\Psi(R)}
\end{equation}
In the variational Monte Carlo simulation the key quantity to be computed is the gradient of the variational energy with respect to the variational parameters. It is given by
\begin{equation}
\nabla E= \langle \nabla \ln \Psi(R) E_{loc}(R) \rangle_{\Psi}-E\langle \nabla \ln \Psi(R) \rangle_{\Psi}
\end{equation}
These quantities can be computed by standard Monte Carlo sampling on the distribution generated by $|\Psi(R)|^2$ with the inverse update trick on Slater determinant. 

In our calculation, instead of optimizing the parameters  $\chi^{\alpha}_{i,j}$ and $\Delta_{i,j}$ in the mean field ansatz $H_{\mathrm{MF}}$, we will optimize the matrix element of the first $2N^{\uparrow}_{e}$ columns of the matrix $\mathbf{U}$. The main advantage of  using the matrix element of $\mathbf{U}$ as the variational parameter is that the computation of the gradient of the wave function now become rather cheap. More specifically, we have
\begin{equation}
\nabla \ln \Psi =\nabla \ln \mathrm{Det}[\bm{\Phi}]=\mathrm{Tr}[\nabla \bm{\Phi}\bm{\Phi}^{-1}]
\end{equation}
here $\bm{\Phi}^{-1}$ is the inverse of matrix $\bm{\Phi}$, which is also used in the inverse update of Slater determinant. Since the variational parameters is just the matrix element of $\mathbf{U}$, the computation of $\nabla \bm{\Phi}$ becomes almost trivial.

Since $\mathbf{U}$ is a complex matrix, there will be $2\times2N\times2N^{\uparrow}_{e}$ real parameters to be optimized. For example, on a kagome cluster with $N=432$ site, the number of variational parameters to be optimized at the $1/9$ plateau will be $N_{v}=663552$(here we have set $2N^{\uparrow}_{e}=N-2M=384$ using the spin inversion symmetry of the KAFH model). Optimizing such large number of parameters in variational Monte Carlo simulation efficiently constitutes a big challenge to the optimization algorithm. In the next section we will introduce an improved optimization algorithm to accomplish such a challenging task.

\section{The finite-depth BFGS algorithm}
The key for a good optimization algorithm is to choose a wise approximation for the Hessen matrix of the system to achieve good balance between numerical stability, numerical efficiency and storage demand. The simplest choice of the identity matrix as the Hessen matrix corresponds to the algorithm of steep descent(SD). While the storage demand of the SD algorithm is the least, it becomes extremely inefficient when there is large fluctuation in the eigenvalues of the true Hessen matrix. In particular, the SD algorithm would get stuck when the optimization enters long and steep valley with very flat bottom in the energy landscape.
           
An important development in the area of variational optimization is the stochastic reconfiguration(SR) algorithm proposed by Sorella\cite{Sorella}. In the SR algorithm, we approximate the Hessen matrix by a positive-definite and Hermitian matrix $\mathbf{S}$ generated from the metric of the variational state in the Hilbert space. More specifically, $\mathbf{S}$ is given by
 \begin{equation}
 \mathbf{S}=\langle \nabla ln\Psi \nabla ln\Psi \rangle-\langle \nabla ln\Psi \rangle\langle \nabla ln\Psi \rangle
 \end{equation}
Since $\nabla ln\Psi$ is also involved in the computation of the local energy, the computation of the matrix element of $\mathbf{S}$ is almost for free. The variational parameter $\bm{\alpha}$ is then updated as follows 
 \begin{equation}
\bm{\alpha}\rightarrow \bm{\alpha}-\delta\ \mathbf{S}^{-1}\nabla E
\end{equation} 
in which $\delta$ is a step length usually chosen by trial and error. The essence of the SR algorithm is to measure distance in Hilbert space rather than in the Euclidean space of the variational parameters, as is done in the SD algorithm. This change can be important when some of the variational parameters become almost redundant for the physical wave function. In such a situation, the fluctuation in the eigenvalue of the Hessen matrix becomes large and the nightmare of the SD algorithm - the long and steep valley with very flat bottom will emerge in the energy landscape. 

The SR algorithm has been applied successfully in many problems. However, we note that since $\mathbf{S}$ is solely determined by the variational state itself, it fails to approximate the true Hessian matrix of the system when the effect of the Hamiltonian matrix element is crucial. In practice, the SR method may still suffer slow convergence or even run away from true minimum. A more serious problem with the SR algorithm is the huge storage demand when we treat a large number of variational parameters, although the computation of the matrix element of $\mathbf{S}$ is almost for free. For example, for the $1/9$ plateau on a kagome cluster with $N=432$ site, $\mathbf{S}$ would be a $663552\times663552$ matrix if the general RVB ansatz proposed in the last section is adopted. This is obviously unrealistic to implement.

In a recent work, we proposed a finite-depth Broyden-Fletcher-Goldfarb-Shanno (BFGS) algorithm for variational optimization. The BFGS method is a quasi-Newton method. It generates an iterative approximation for the inverse of the Hessian matrix\cite{Nocedal} from the gradient of variational energy. The approximate inverse Hessian matrix is updated as follows
\begin{equation}
\mathbf{B}_{k+1}=\left(\mathbf{I}-\frac{\mathbf{s}_{k}\mathbf{y}^{T}_{k}}{\mathbf{y}^{T}_{k}\mathbf{s}_{k} } \right)\mathbf{B}_{k}\left(\mathbf{I}-\frac{\mathbf{y}_{k}\mathbf{s}^{T}_{k}}{\mathbf{y}^{T}_{k}\mathbf{s}_{k} } \right)+\frac{\mathbf{s}_{k}\mathbf{s}^{T}_{k}}{\mathbf{y}^{T}_{k}\mathbf{s}_{k} }
\end{equation}
in which $k=0,1,2.....$
\begin{eqnarray}
\mathbf{s}_{k}&=&\bm{\alpha}_{k+1}-\bm{\alpha}_{k}\nonumber\\
\mathbf{y}_{k}&=&\nabla E_{k+1}-\nabla E_{k}
\end{eqnarray}  
are the difference between successive variational parameters and energy gradients. $\bm{\alpha}_{k=0}$ is the initial guess of the variational parameters and $\nabla E_{k=0}$ is the energy gradient at the starting point. The Hessian matrix is initially set to be the identity matrix, namely $\mathbf{B}_{k=0}=\mathbf{I}$. Using such an iterative approximation on the inverse of the Hessian matrix, the variational parameters are updated as follows
\begin{equation}
\bm{\alpha}_{k+1}=\bm{\alpha}_{k}+\delta\ \mathbf{B}_{k} \nabla E_{k}
\end{equation}  
Here $\delta$ is the step length chosen by trial and error. 

In practice, we restart the BFGS iteration every $K$ steps.  We call such an algorithm $K-$depth BFGS algorithm. The advantage of the finite-depth BFGS algorithm lies in the fact that we do not need to store large matrix any more. In fact, since all we need in the $K-$depth BFGS iteration is the product of the matrix $\mathbf{B}_{k}$ with vector $\nabla E_{k}$ or $\mathbf{y}_{k}$ and the inner product of vector $\mathbf{s}_{k}$ and $\mathbf{y}_{k}$, we only need to store the $2K$ vectors $\mathbf{s}_{k}$ and $\mathbf{y}_{k}$. The needed matrix-vector and vector inner product can then be computed recursively using these $2K$ vectors.

\section{The magnetization curve of the spin-1/2 KAFH and the nature of the plateau states}
Since the total magnetization is a conserved quantity of the system, we have optimized the RVB wave function with the finite-depth BFGS algorithm for each value of the magnetization. The calculation is done on kagome cluster with both $N=108$ and $N=432$ sites as illustrated in Fig.1. These kagome clusters all have the $N=L\times L\times 3$ geometry, with periodic boundary condition imposed on both the $\mathbf{a}_{1}$ and the $\mathbf{a}_{2}$ direction. Here $L$ is the number of kagome unit cell in the $\mathbf{a}_{1}$ and the $\mathbf{a}_{2}$ direction. We note that these clusters contain an even number of kagome unit cells so that one can reach the true ground state in the absence of the external field. At the same time, since $L$ is a multiple of $3$, these clusters can all accommodate the $\sqrt{3}\times\sqrt{3}$ VBC pattern of the $1/3$, $5/9$ and the $7/9$ plateau state. 

\begin{figure}
\includegraphics[width=8.5cm]{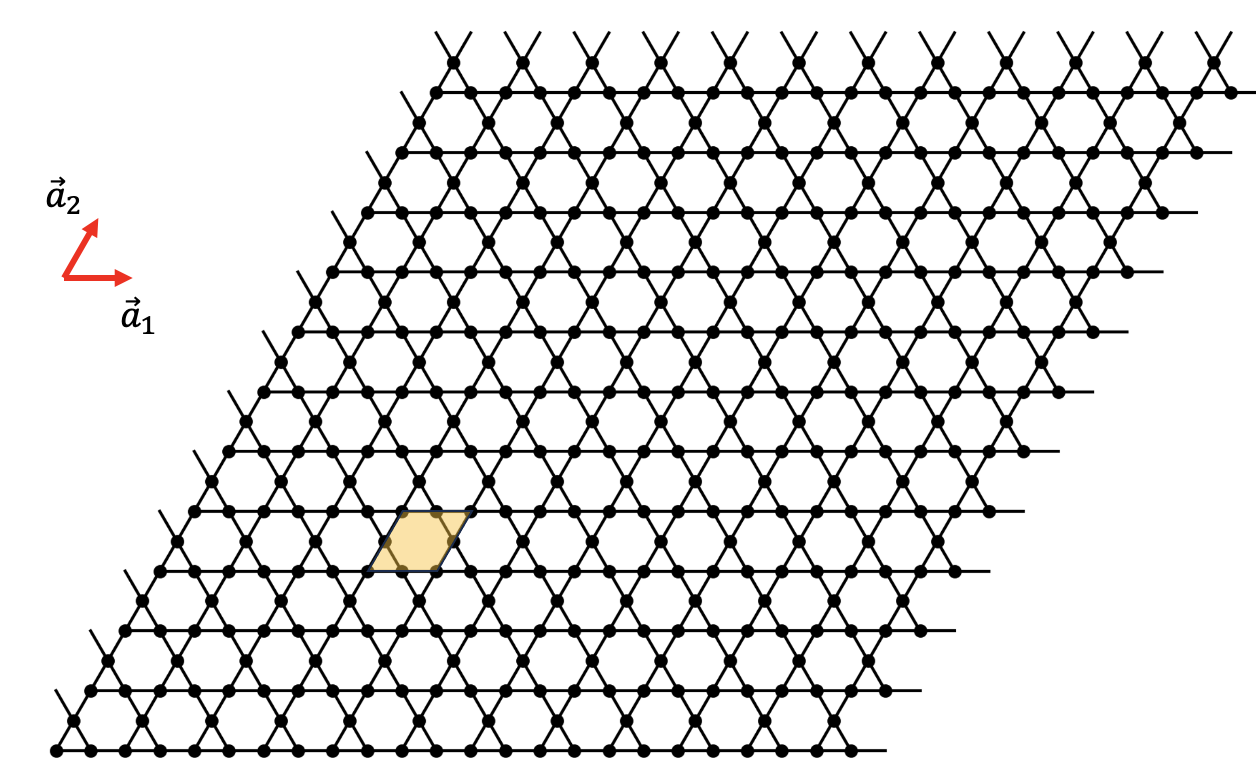}
\caption{Illustration of the kagome cluster on which our variational calculation is done. Periodic boundary condition is assumed along both the $\mathbf{a}_{1}$ and the $\mathbf{a}_{2}$ direction. Here $\mathbf{a}_{1}$ and $\mathbf{a}_{2}$ are the basis vectors of the kagome lattice. The unit cell of the kagome lattice is illustrated by the brown shaded parallelogram. }
\end{figure}

\subsection{The ground state energy and the magnetization curve}
\subsubsection{The ground state energy}
Our calculation of the magnetization curve is carried out using canonical ensemble with conserved total magnetization $M$. To proceed, we perform variational optimization of the ground state energy of the spin-$\frac{1}{2}$ KAFH in the absence of the external field, denoted here as $E(M)$, for each given value of $M$. The energy of the system in the presence of the external field is then given by
\begin{equation}
E(M,B)=E(M)-BM
\end{equation} 
The magnetization of the system at external field $B$ is then determined by minimizing $E(M,B)$ with respect to $M$ at fixed value of $B$, or from the following condition
\begin{equation}
\frac{\partial E(M,B)}{\partial M}\Big|_{B}=0
\end{equation} 
In practice, we overlay the $E(M,B)$ curves as a function of $B$ calculated at all different values of $M$. We then find the lower boundary of these group of curves, which is nothing but the optimized ground state energy of the system in the external field. This is illustrated for the $N=432$ kagome cluster in Fig.2a in the whole range of magnetization, which contains $217$ possible values in the range of $M\in[0,216]$. 

As is shown in Fig.2b, the ground state energy obtained from our unrestricted variational optimization is significantly lower than that reported in previous works. This is especially clear in $1/9$ plateau regime where a $Z_{3}$ chiral spin liquid state has been claimed. More specifically, the energy of the $Z_{3}$ chiral spin liquid state obtained by previous variational study is $E(M)\approx-0.4116J$/site\cite{JXLi}. Mysteriously, this is almost identical to the ground state energy found by previous tensor network simulation, namely $E(M)\approx-0.4111J$/site\cite{Fang}, in which a $\sqrt{3}\times\sqrt{3}$ VBC state is claimed. The energy obtained from our unrestricted optimization is much lower, reaching a value as low as $E(M)\approx-0.4184J$/site.

\begin{figure}
\includegraphics[width=8cm]{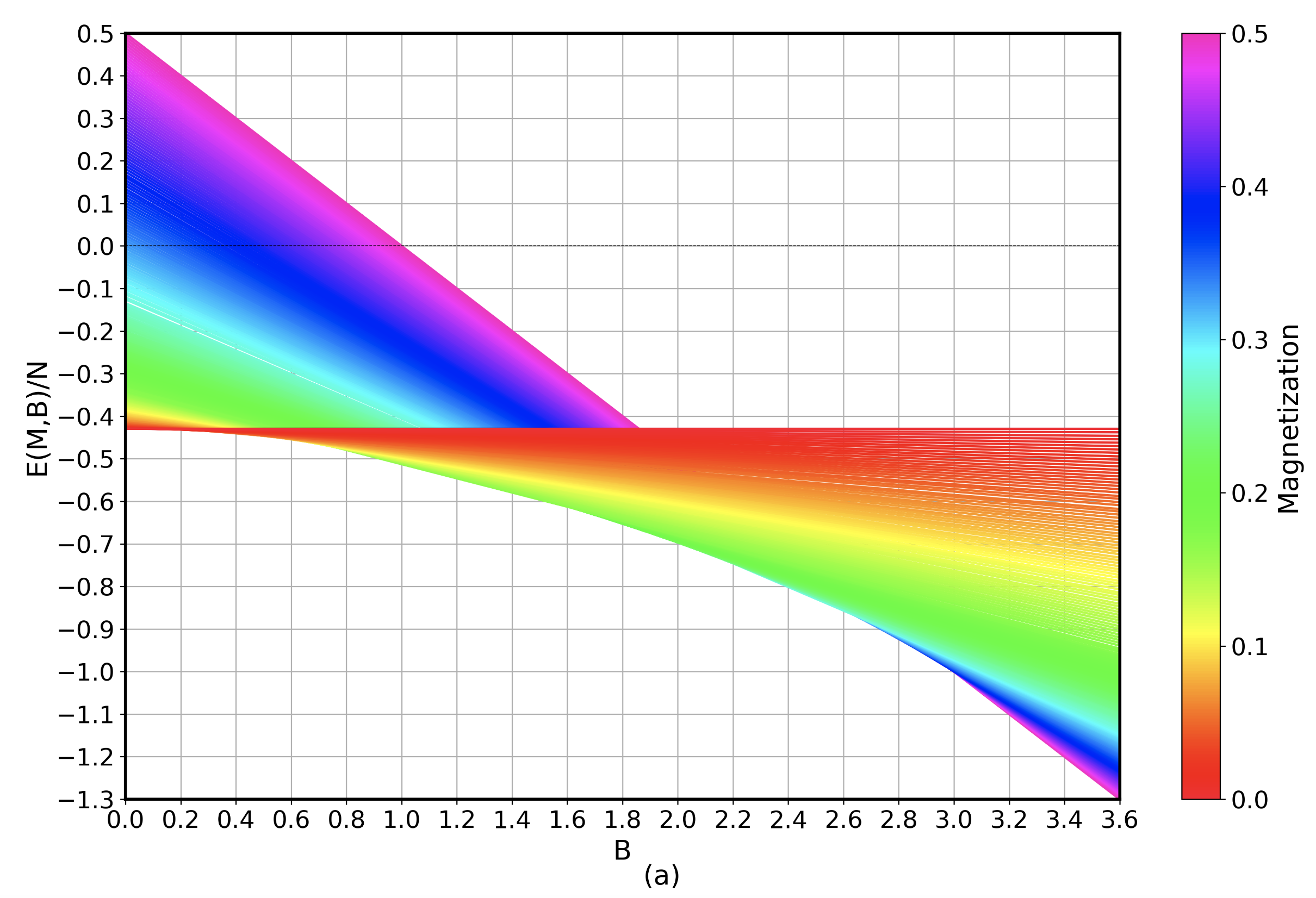}
\includegraphics[width=8cm]{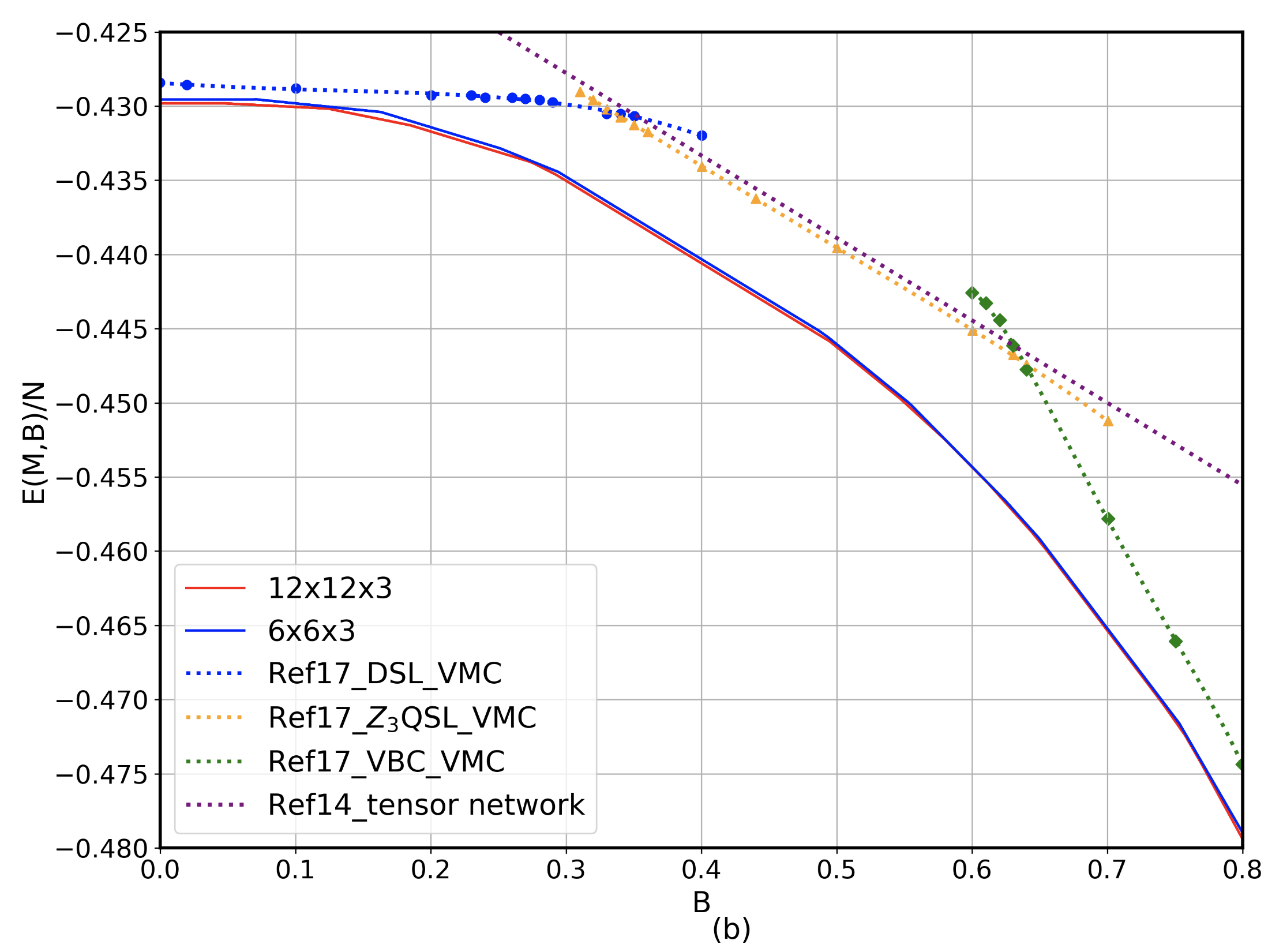}
\caption{(a)Overlay of the $E(M,B)$ curves as a function of $B$ calculated at all possible value of $M$ for a $N=432$ kagome cluster. The value of the magnetization is shown in color scale. The lower boundary of these group of curves gives the ground state energy of the system in the external field. (b)The ground state energy as a function of $B$ calculated in this work compared with several previous theoretical results. Here the energy obtained by tensor network simulation at the $1/9$ plateau is extended into a line with slope $-\frac{1}{18}$ for ease of comparison.}
\end{figure}

We note that the energy calculated on the $N=432$ cluster is very close to that calculated on the $N=108$ cluster, implying that the $N=432$ cluster is already large enough to reach conclusion valid in the thermodynamic limit. We also note that the energy curve of the $N=432$ cluster lies slightly below the energy curve of the $N=108$ cluster. This simply results from the fact we have a finer resolution in the magnetization on the $N=432$ cluster.

\subsubsection{The magnetization curve}
The magnetization curve determined by the above procedure for both the $N=108$ and the $N=432$ cluster is shown in Fig.3. Clearly the finite size effect is still very strong on such small clusters. Nevertheless, the plateau at $M=1/9,1/3,5/9$ and $7/9$ can all be clearly resolved. Moreover, the plateau width calculated on the $N=108$ and the $N=432$ cluster are almost identical. The detailed field range of the four magnetization plateaus are summarized in Tab.I.

\begin{figure}
\includegraphics[width=8cm]{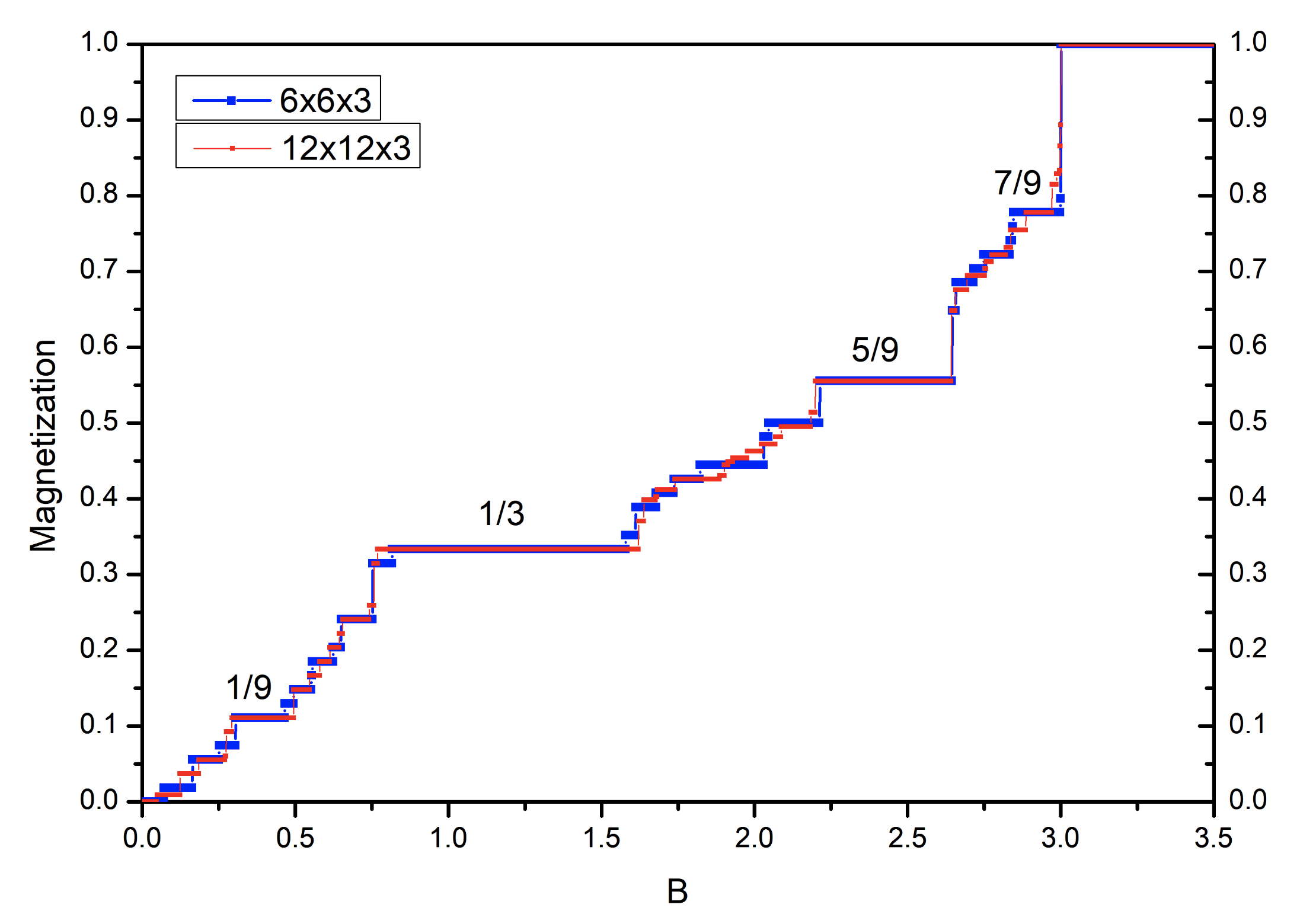}
\caption{Comparison of the magnetization curve of the spin-1/2 KAFH on the $N=6\times6\times3=108$ and the $N=12\times12\times3=432$ kagome cluster in the full magnetization range.}
\end{figure}

\begin{table}
\begin{tabular}{|c|c|c|c|c|}
\hline
\diagbox{N}{plateau} & 1/9&1/3& 5/9 & 7/9\\
\hline
108& [0.31,0.47] &  [0.82,1.58]  & [2.21,2.65] & [2.85,3] \\
\hline
432& [0.29,0.49] &  [0.77,1.62] &  [2.2,2.64] & [2.89,2.97] \\
\hline
\end{tabular}
\caption{\label{tab:test}Comparison of the field range of the four magnetization plateaus obtained on the $N=108$ and the $N=432$ cluster.} 
\end{table}

Our large scale unrestricted variational study does not support the existence of so broad a $1/9$ plateau as claimed in Ref.\onlinecite{JXLi}. Moreover, as will be detailed below, we find that the $1/9$ plateau state features a novel $3\times3$ VBC pattern with a windmill-shaped motif, rather than the $Z_{3}$ chiral spin liquid state claimed in Ref.\onlinecite{JXLi}. Since  our general RVB ansatz contains the variational ansatz adopted in Ref.\onlinecite{JXLi} as a subset, we think the claimed stability of the $Z_{3}$ chiral spin liquid state is an artifact resulted from the variational bias in the NN-$U(1)$ ansatz adopted in Ref.\onlinecite{JXLi}.

\subsection{The nature of the magnetization plateaus}
\subsubsection{The nature of the $1/3$ plateau state}
\begin{figure}
\includegraphics[width=8cm]{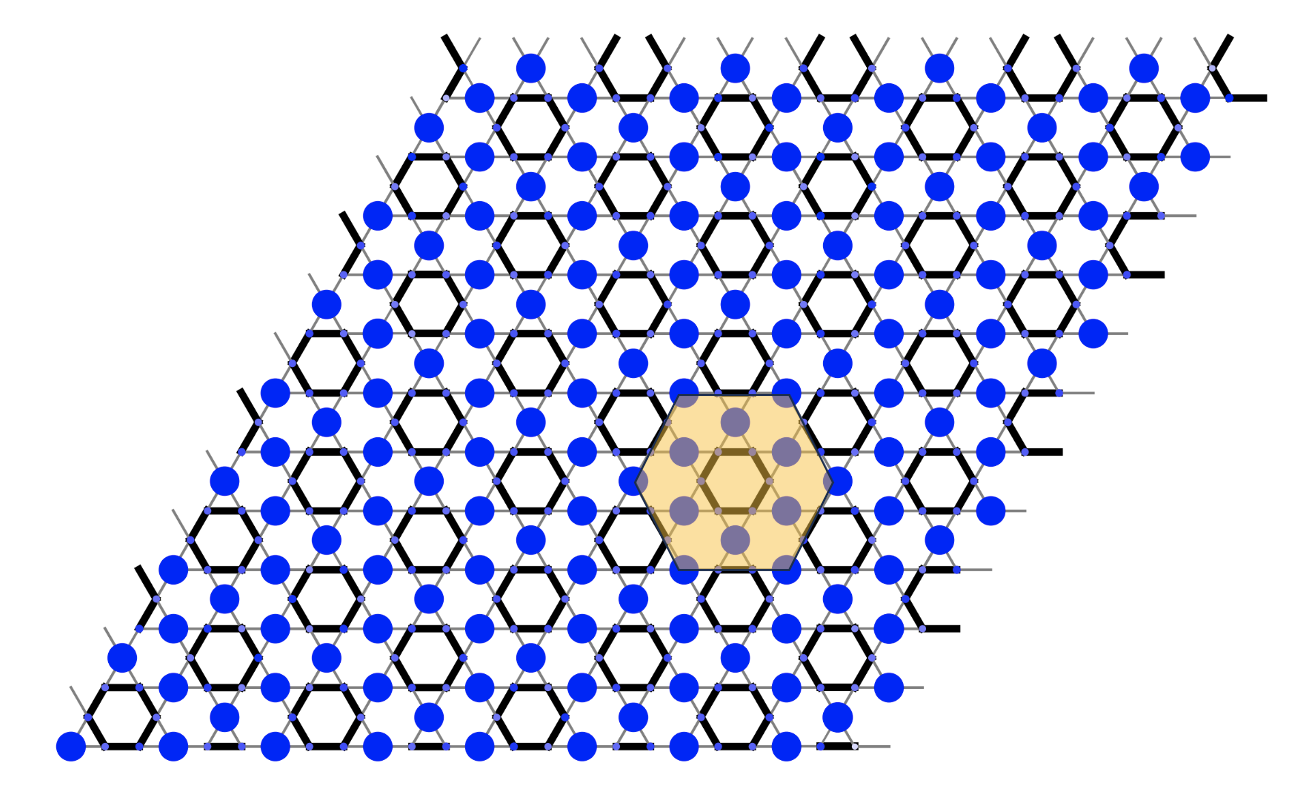}
\caption{The $\sqrt{3}\times\sqrt{3}$ VBC pattern in the $1/3$ plateau state as obtained by our variational optimization. Here we use the size of the solid dots to represent the magnitude of the magnetization on a give lattice site and the thickness of the bond to represent the local spin correlation between nearest-neighboring sites. The brown shaded hexagon marks the David-star-shaped motif of the $\sqrt{3}\times\sqrt{3}$ VBC pattern.}
\end{figure}

The ground state at the $1/3$ plateau is characterized by the well known $\sqrt{3}\times\sqrt{3}$ VBC pattern with a David-star-shaped motif as illustrated in Fig.4. The unit cell of the David-star VBC pattern contains 9 sites, among which 6 sites reside on the inner hexagonal ring of the David-star and the remaining 3 sites reside on its vertices. The magnetization is mainly contributed by the 3 sites on the vertices of the David star. Such a VBC pattern is well reproduced by our variational optimization. In Fig.4 we plot the distribution of the magnetization and the strength of local spin correlation between nearest neighboring sites in the optimized ground state at the $1/3$ plateau. In our computation, we have frequently encountered solution with domain wall defect in the VBC pattern. To eliminate such imperfection, we have performed induced optimization in which we have strengthened the exchange coupling on the hexagonal rings of the David-star motif. The solution of the induced optimization is then relaxed to find the true ground state of the spin-$1/2$ KAFH at the $1/3$ plateau. We find that the domain wall imperfection can be successfully removed by such an induction-relaxation procedure and the energy of the relaxed solution is lower than that obtained by direct optimization.

More specifically, we find that the magnetization on the vertices of the David star is $\langle \mathbf{s}^{z}_{i}\rangle\approx 0.435$. This is much larger than that on the inner hexagonal ring, which is $\langle \mathbf{s}^{z}_{i}\rangle\approx 0.035$. The local spin correlation on the inner hexagonal ring of the David star is $\langle \mathbf{s}_{i}\cdot\mathbf{s}_{j}\rangle\approx -0.4024$. This is much stronger than that between the vertices of the David star and the sites on the inner hexagonal ring of the David star, which is $\langle \mathbf{s}_{i}\cdot\mathbf{s}_{j}\rangle\approx -0.0584$. We note that the local magnetization on both the vertices and the inner hexagonal ring of the David star are polarized in the same direction as the external field in the $1/3$ plateau state.  

\subsubsection{The nature of the $5/9$ and the $7/9$ plateau state}
We find that the same $\sqrt{3}\times\sqrt{3}$ VBC pattern with a David-star motif also appears in the $5/9$ and the $7/9$ plateau state. The magnetization is still mainly contributed by the 3 vertices of the David-star motif. In particular, in the $7/9$ plateau state, the magnetization on the vertices is almost saturated. In addition, the local magnetization is always polarized in the same direction as the external field. As we will see below, this is qualitatively different from what we find in the more exotic $1/9$ plateau state, in which local magnetization can be polarized in the opposite direction of the external field. This implies that the system at the $1/3$, $5/9$ and the $7/9$ plateau is less frustrated than system at the $1/9$ plateau. A comparison of the detailed distribution in the local magnetization and the local spin correlation for the $1/3$, $5/9$ and the $7/9$ plateau state are provided in Tab.II and Tab.III.

\begin{table}
\begin{tabular}{|c|c|c|c|}
\hline
\diagbox{site position}{plateau} & 1/3& 5/9 & 7/9\\
\hline
vertex& 0.435 & 0.476 &1/2\\
\hline
hexagonal ring& 0.035& 0.179& 1/3\\
\hline
\end{tabular}
\caption{\label{tab:test}The local magnetization on the vertices and the inner hexagonal ring of the David-star-shaped motif for the $1/3$, $5/9$ and the $7/9$ plateau.} 
\end{table}

\begin{table}
\begin{tabular}{|c|c|c|c|}
\hline
\diagbox{bond type}{plateau} & 1/3& 5/9 & 7/9\\
\hline
vertex-ring& -0.0584 & -0.061 &0.1663\\
\hline
within ring& -0.4024& -0.3238& -0.0815\\
\hline
\end{tabular}
\caption{\label{tab:test}The local spin correlation between nearest-neighboring sites on the David-star-shaped motif for the $1/3$, $5/9$ and the $7/9$ plateau.} 
\end{table}

\subsubsection{The nature of the $1/9$ plateau state}
The nature of the $1/9$ plateau state is much more subtle than the $1/3$ plateau state discussed above. This is consistent with the fact that the $1/9$ plateau is much narrower than the $1/3$ plateau. From intensive refined optimization, we find that the $1/9$ plateau state is characterized by a 
$3\times3$ VBC pattern with a windmill-shaped motif as illustrated in Fig.5. The unit cell of the windmill-shaped VBC pattern contains 27 sites, among which 6 sites reside on the inner hexagonal ring of the windmill,12 sites reside on the 6 wings of the windmill and the remaining 9 sites reside on the 3 dumbbells on or across the boundary of the parallelogram-shaped unit cell. We find that the energy gain to form such a VBC pattern is much shallower than that of the David-star VBC pattern realized at the $1/3$ plateau. As a result, the windmill VBC pattern is much more vulnerable than the David-star VBC pattern and domain wall defect occurs much more frequently at the $1/9$ plateau. To eliminate such imperfection, we have also carried out induced optimization. We find the windmill pattern survives the relaxation from such induced optimization with an energy lower than that obtained from direct optimization.

We note that several other VBC patterns have been proposed in the literature for the $1/9$ plateau state. In particular, the type-1 VBC pattern proposed in Ref.\onlinecite{Morita} is equivalent to the VBC pattern claimed in Ref.\onlinecite{Fang} and both are thought to be the most probable VBC pattern at the $1/9$ plateau. Unlike the windmill-shaped VBC pattern found here, the type-1 VBC pattern proposed in Ref.\onlinecite{Morita} features a $\sqrt{3}\times\sqrt{3}$ periodicity and a smaller unit cell with 9 sites. We have also carried out induced optimization for this VBC pattern but find it does not survive after relaxation. In addition, we note that the windmill VBC pattern has a higher symmetry than the type-1 VBC pattern proposed in Ref.\onlinecite{Morita}. More specifically, the local spin correlation in the windmill VBC pattern has a nice six-fold rotational symmetry. On the other hand, the local spin correlation in the type-1 VBC pattern only features a much weaker mirror symmetry about a single reflection plane. At the same time, the energy of the windmill VBC pattern is significantly lower than that of the type-1 VBC pattern proposed in Ref.\onlinecite{Morita}.  As we stated above, the energy of the windmill VBC state is $-0.4184J$/site. This is much lower than the energy of the type-1 VBC pattern found from tensor network simulation, which is $-0.4111$/site. 

\begin{figure}
\includegraphics[width=8cm]{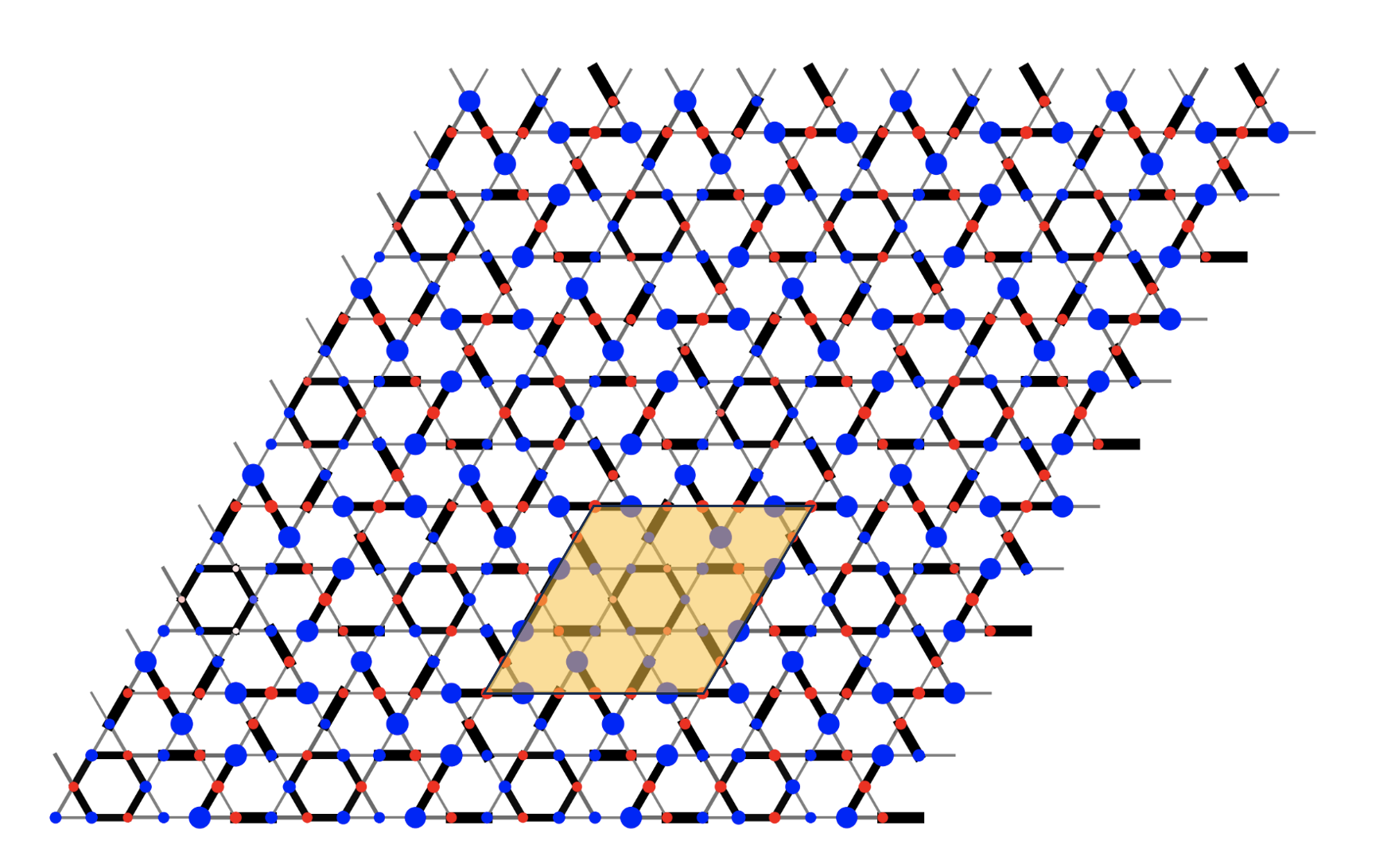}
\includegraphics[width=8cm]{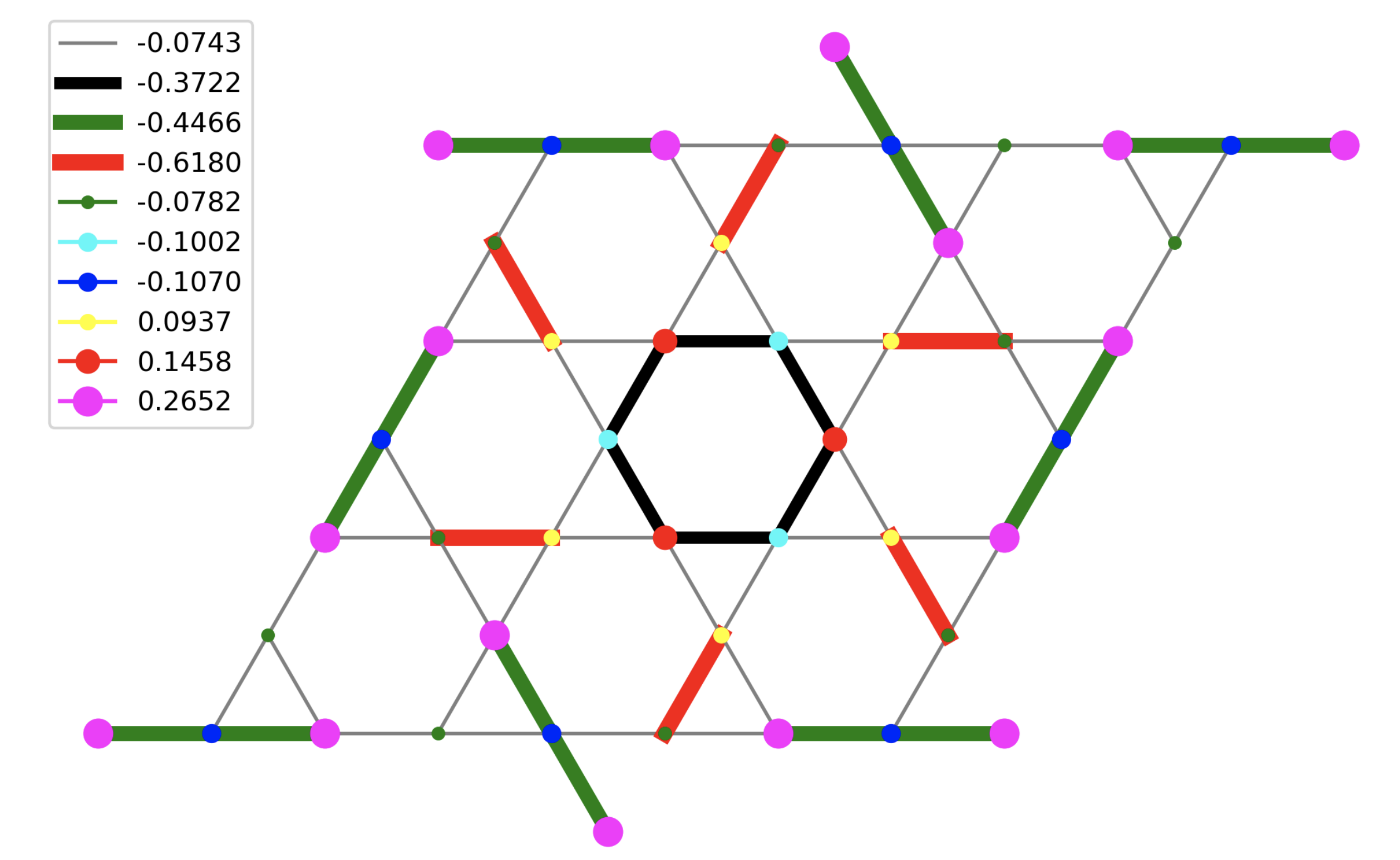}
\caption{Upper panel: The $3\times3$ VBC pattern in the $1/9$ plateau state as obtained by our intensive refined variational optimization. Here we use the size of the solid dots to represent the magnitude of the magnetization on a give lattice site and the thickness of the bonds to represent the strength of local spin correlation between nearest-neighboring sites. The brown shaded parallelogram marks the unit cell of the windmill-shaped VBC pattern. Lower panel: An enlarged view of the windmill-shaped motif of the VBC pattern shown in the upper panel with detailed distribution of local magnetization and local spin correlation indicated.}
\end{figure}

A major difference between the windmill VBC pattern realized at the $1/9$ plateau and the David-star VBC pattern realized at the $1/3$ plateau is that the spatial modulation of magnetization at the $1/9$ plateau is so strong that even its polarization can be reversed. More specifically, we find that there are three groups of site in the windmill VBC pattern that contribute differently to the magnetization. The magnetization is mainly contributed by the 9 sites residing on the 3 dumbbells on or across the boundary of the parallelogram-shaped unit cell. The magnetization reversion mainly occurs on the center sites of the 3 dumbbells, which is surrounded by two end sites with strong positive magnetization. The 6 sites on the inner hexagonal ring and the 12 sites on the 6 surrounding wings of the windmill collectively have little contribution to the magnetization. On the other hand, the local spin correlation is much stronger on the inner hexagonal ring and the surrounding wings than those on the dumbbells and those between sites belonging to different groups. The detailed distribution of the local magnetization and the local spin correlation can be found in the Fig.5. We note that reversal of polarization in the local magnetization is not rare for strongly frustrated magnets. The most important example in this respect is the up-up-down state at the $1/3$ plateau of the triangular antiferromagnet of either classical or quantum type, in which $1/3$ of the lattice site are polarized oppositely with respect to the external field\cite{Kawamura}. 
 
\subsubsection{The nature of the ground state for general magnetization between the plateaus}
We find that the magnetization of the system is never homogeneous below the saturation field. For a general magnetization away from the plateaus, we find that the local magnetization in general features incommensurate spatial modulation. However, we find that the ground state energy of the system is rather insensitive to the exact position of the modulation wave vector.  It is thus rather difficult to determine the exact nature of the symmetry breaking pattern in the ground state at a general magnetization. Nevertheless, we find that the strength of the spatial inhomogeneity in the magnetization evolves smoothly with the total magnetization $M$. To illustrate this point, we have calculated the standard deviation of the local magnetization. It is defined by
\begin{equation}
\sigma_{m}=\sqrt{\frac{1}{N}\sum_{i=1}^{N}(m_{i}-m)^{2}}
\end{equation}
here $m=M/N$ is the average magnetization on a given site.  The result of $\sigma_{m}$ as a function of $m$ is plotted in Fig.6. For small $m$, we find that
\begin{equation}
\sigma_{m}\propto \sqrt{m}
\end{equation}   
Thus the relative inhomogeneity in the magnetization, namely $\sigma_{m}/m$, diverges in the zero field limit. The spin-$\frac{1}{2}$ KAFH is thus very different from a conventional paramagnet, for which a spatially uniform magnetization is expected in a weak uniform field. We note that a similar strong inhomogeneity in the weak field magnetic response is predicted for the spin-$\frac{1}{2}$ KAFH when there is a single local impurity in the system\cite{Yang2}. The strong tendency toward spatial inhomogeneous magnetic response of the system, which is counterintuitive from the perspective of conventional weakly correlated paramagnet, should be attributed to the strongly frustrated nature of the spin-$\frac{1}{2}$ KAFH. In the more extended range of $m$, the relative inhomogeneity in the magnetization $\sigma_{m}/m$ is found to decrease monotonically with $m$, until it vanishes at the saturation field. We also note that there are local minima in $\sigma_{m}$ around the $1/3$ and $7/9$ plateau, implying that the system are less frustrated at these plateaus as compared to nearby magnetization.

\begin{figure}
\includegraphics[width=8cm]{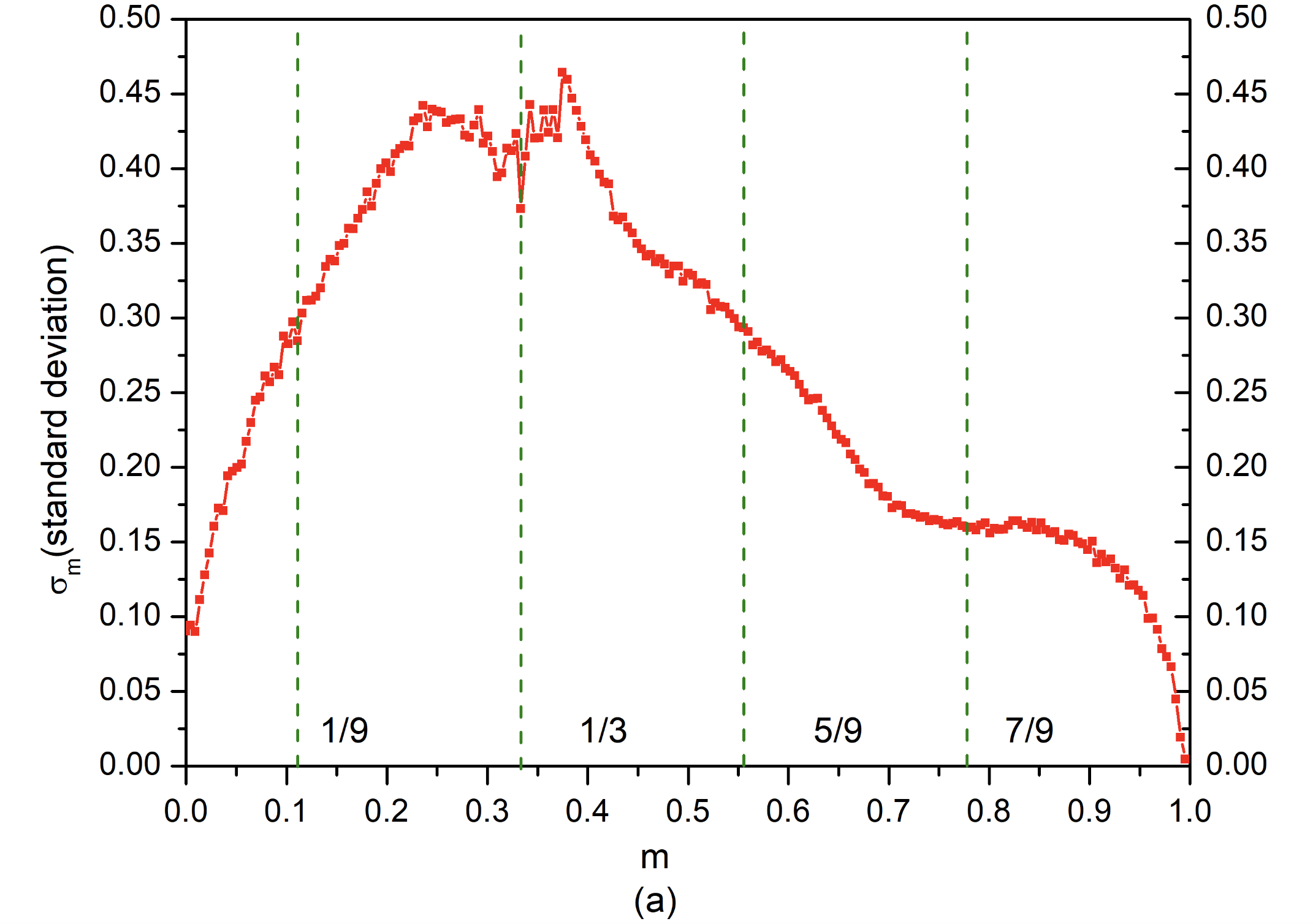}
\includegraphics[width=8cm]{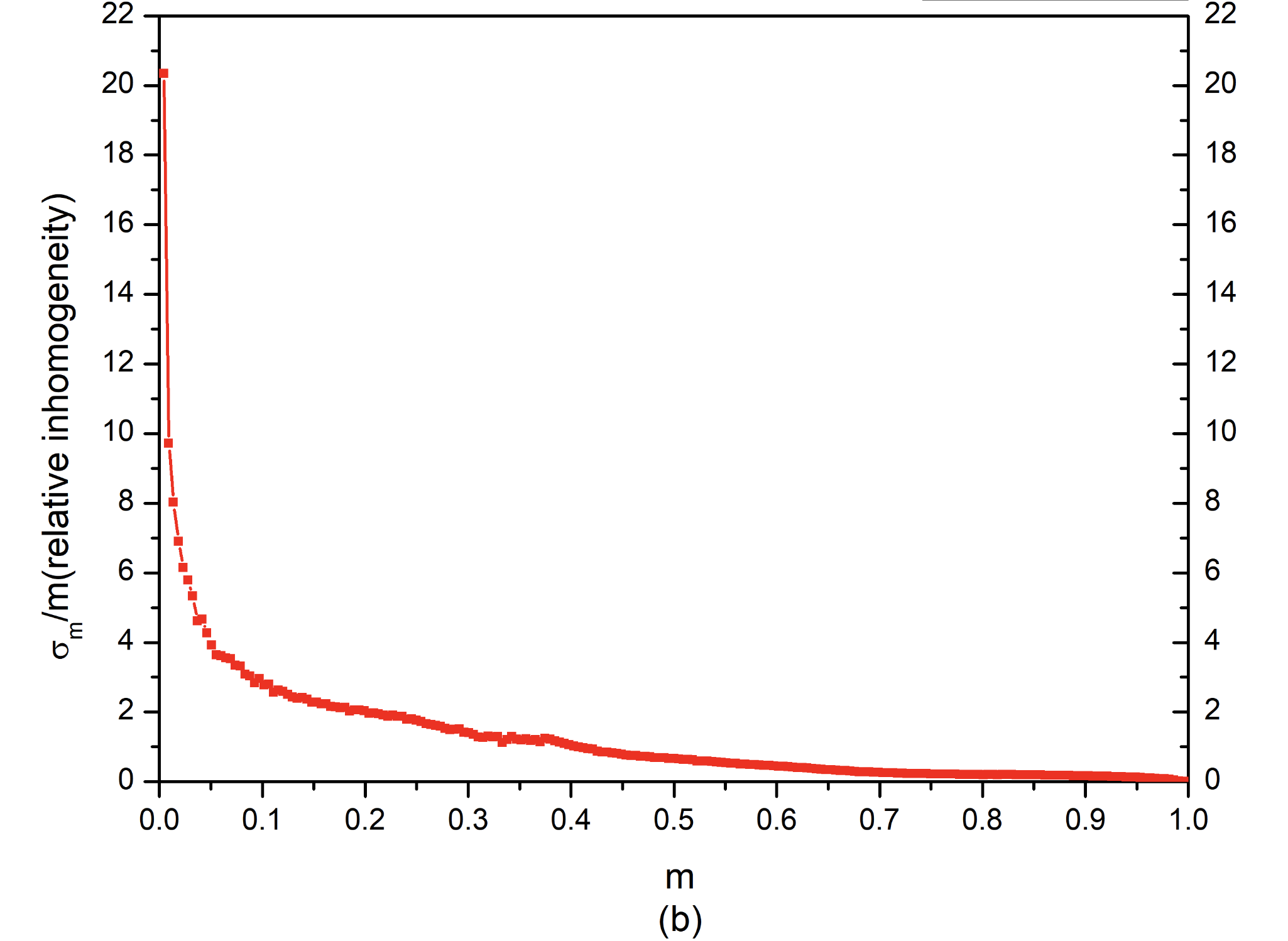}
\caption{(a)The standard deviation in the local magnetization $\sigma_{m}$ as a function of the average magnetization $m$. (b)The relative inhomogeneity in the local magnetization, namely $\sigma_{m}/m$ as a function of the average magnetization $m$.}
\end{figure}

\section{Conclusions and Discussions}
The origin of the $1/9$ plateau in the magnetization curve of the spin-$\frac{1}{2}$ KAFH remains elusive after a decade of intense theoretical investigation. The interest in this problem is recently refuelled by experimental claims of the observation of the $1/9$ plateau in the kagome spin liquid candidate material YCOB. The main theoretical challenge to resolve the controversy about this issue is to reduce the bias of the method used to describe the magnetization process of such a strongly frustrated quantum magnet, for which intricate competition between different symmetry breaking patterns is generally expected. For this purpose, we have proposed in this work the most general RVB ansatz that is consistent with the spin symmetry of the system. The proposed RVB ansatz can describe all kinds of spatial symmetry breaking pattern on equal footing since no spatial symmetry is assumed on it. To optimize such a variational ansatz, which contains a huge number of variational parameters, we have proposed a new optimization algorithm, namely the finite-depth BFGS algorithm. The new algorithm can achieve good balance between numerical efficiency, numerical stability and storage demand.

We have applied such an advanced optimization algorithm to the general RVB ansatz we have proposed and mapped out the magnetization curve of the spin-$\frac{1}{2}$ KAFH. We find that the variational energy obtained from such a general RVB ansatz is significantly lower than that obtained from previous variational study using nearest-neighboring $U(1)$ RVB ansatz and that from tensor network simulations. The energy difference is especially large in the $1/9$ plateau regime in which a robust $Z_{3}$ chiral spin liquid state has been claimed by previous variational study. We find that the $1/9$ plateau features a windmill-shaped VBC pattern with a $3\times3$ enlarged unit cell. Such a VBC pattern is totally different from the VBC pattern anticipated before from tensor network simulation and ED analysis, which has a $\sqrt{3}\times\sqrt{3}$ unit cell. The windmill-shaped VBC pattern features nice six-fold rotational symmetry in local spin correlation and exhibits strong spatial modulation in the local magnetization. The modulation in the magnetization is so strong that even its polarization can be reversed. 
 
Our general RVB ansatz can also well reproduce the other three magnetization plateaus of the spin-$\frac{1}{2}$ KAFH. Unlike the $1/9$ plateau, we find that the $1/3$, $5/9$ and $7/9$ plateau all feature the well known $\sqrt{3}\times\sqrt{3}$ VBC pattern with a David-star-shaped motif. In addition, we find that there is no reversion in the polarization of local magnetization at these more conventional plateaus. 

Besides the magnetization plateaus, we have also studied the structure of the ground state at general magnetization in between the plateaus.  We find that the magnetization of the system is always strongly inhomogeneous. However, it is much harder to determine the exact nature of the spatial symmetry breaking pattern at such general magnetization. Nevertheless, we find that the strength of spatial inhomogeneity in the local magnetization evolves smoothly with the magnetic field. More specifically, we find that $\sigma_{m}/m$ decreases almost monotonically with increasing field, until it vanishes at the saturation field. In the weak field regime , we find that the relative inhomogeneity diverges as $1/\sqrt{m}$. Such strong spatial inhomogeneity in the local magnetization should obviously be attributed to the strongly frustrated nature of the spin-$\frac{1}{2}$ KAFH.  

Finally, we discuss the possible relevance of this work to the recent experimental observations on the kagome spin liquid candidate material YCOB. According to this study, all magnetization plateaus of the spin-$\frac{1}{2}$ KAFH are translational symmetry breaking VBC phase, although the VBC pattern at the $1/9$ plateau has a more complicated windmill-shaped motif and a larger unit cell containing 27 sites. We think that the experimentally claimed $1/9$ plateau in YCOB should correspond to such a VBC phase, rather than the more exotic $Z_{3}$ chiral spin liquid phase. As we have shown in this paper, the magnetic response of the spin-$\frac{1}{2}$ KAFH is always characterized by strong spatial inhomogeneity. This prediction should be checked by local measurement such as nuclear magnetic resonance. Such strong spatial modulation in the magnetization may also be the origin of the claimed quantum oscillation in the torque measurement of YCOB. More specifically, metastable plateaus may occur between the major plateaus as a result of the proximity of the modulation wave vector of the local magnetization to some commensurate values. Moreover, as the exchange couplings in the YCOB material may be disordered, the interplay between such disorder effect and the intrinsic tendency of the spin-$\frac{1}{2}$ KAFH toward spatially inhomogeneous magnetic response is an interesting issue to be answered. These issues are clearly within the reach of our approach. We left them to future investigations.

\begin{acknowledgments}
We acknowledge the support from the National Natural Science Foundation of China(Grant No.12274457).
\end{acknowledgments}


\begin{thebibliography}{}

\bibitem{Shender}E. F. Shender, Antiferromagnetic garnets with fluctuationally interacting sublattices, Sov. Phys. JETP, \textbf{56}, 178(1982).
\bibitem{Kawamura}H.  Kawamura, Spin wave analysis of the antiferromagnetic plane rotator model on the triangular lattice - symmetry breaking in a magnetic field, J.Phys.Soc.Jpn. \textbf{53}, 2452(1984).
\bibitem{Chubukov}A. V.  Chubukov and D. I. Golosov, Quantum theory of an antiferromagneton a triangular lattice in a magnetic field, J. Phys.: Condens.Matter \textbf{3} 69( 1991).
\bibitem{Hida}K. Hida, Magnetization process of the S=1  and 1/2 uniform and distorted kagome Heisenberg antiferromagnets. J. Phys. Soc. Jpn \textbf{70}, 3673(2001).
\bibitem{Richter1}Honecker, A., Schulenburg, J.  Richter, J. Magnetization plateaus in frustrated antiferromagnetic quantum spin models. J. Phys. Condens. Matter \textbf{16}, S749(2004).
\bibitem{Zhitomirsky}M.E. Zhitomirsky and H. Tsunetsugu, Exact low temperature behavior of a kagome antiferromagnet at high fields, Phys. Rev. B \textbf{70}, 100403(R) (2004).
\bibitem{Sakai}H. Nakano and T. Sakai, Magnetization process of kagome lattice Heisenberg antiferromagnet, J. Phys. Soc. Jpn. \textbf{79}, 053707 (2010).
\bibitem{Nakano}T. Sakai and H. Nakano, Critical magnetization behavior of the triangular and kagome lattice quantum antiferromagnets, Phys. Rev. B \textbf{83}, 100405(R) (2011).
\bibitem{Hotta1}S. Nishimoto, N. Shibata, and C. Hotta, Controlling frustrated liquids and solids with an applied field in a kagome Heisenberg antiferromagnet, Nat. Commun. \textbf{4}, 2287 (2013).
\bibitem{Richter2}S. Capponi, O. Derzhko, A. Honecker, A. M. Läuchli, and J. Richter, Numerical study of magnetization plateaus in the spin-1/2 kagome Heisenberg antiferromagnet, Phys. Rev. B \textbf{88}, 144416 (2013).
\bibitem{Poiblanc}T. Picot, M. Ziegler, R. Orus, and D. Poilblanc, Spin-S kagome quantum antiferromagnets in a field with tensor networks, Phys. Rev. B \textbf{93}, 060407(R) (2016).
\bibitem{Richter3}J. Schnack, J. Schulenburg, A. Honecker, and J. Richter, Magnon crystallization in the kagome lattice antiferromagnet, Phys. Rev. Lett. \textbf{125}, 117207 (2020).
\bibitem{Oshikawa}M. Oshikawa, M. Yamanaka and I. Affleck, Magnetization plateaus in spin chains; ‘‘Haldane gap’’ for half-integer spins. Phys. Rev. Lett. \textbf{78}, 1984 (1997).
\bibitem{Fang}D. Z. Fang, N. Xi, S.-J. Ran, and G. Su, Nature of the 1/9-magnetization plateau in the spin-1/2 kagome Heisenberg antiferromagnet, Phys. Rev. B \textbf{107}, L220401 (2023).
\bibitem{Hotta2}C. Hotta and N. Shibata, Grand canonical finite-size numerical approaches: A route to measuring bulk properties in an applied field, Phys. Rev. B \textbf{86}, 041108 (2012).
\bibitem{Hotta3}C. Hotta,  S. Nishimoto and . Shibata, Grand canonical finite size numerical approaches in one and two dimensions: real space energy renormalization and edge state generation. Phys. Rev. B \textbf{87}, 115128 (2013).
\bibitem{JXLi}L.-W. He, S.-L. Yu, and J.-X. Li, Variational monte carlo study of the 1/9-magnetization plateau in kagome antiferromagnets, Phys. Rev. Lett. \textbf{133}, 096501 (2024).
\bibitem{Morita} K. Morita, Valence bond crystal ground state of the 1/9 magnetization plateau in the spin-1/2 kagome lattice, J. Phys. Soc. Jpn. \textbf{93}, 123706 (2024)


\bibitem{XHChen}X.-H. Chen, Y.-X. Huang, Y. Pan, and J.-X. Mi, Quantum spin liquid candidate YCu$_{3}$(OH)$_{6}$Br$_{2}$[Br$_{x}$(OH)$_{1-x}$]($x \approx0.51$): with an almost perfect kagome layer, J. Magn. Magn. Mater. \textbf{512}, 167066 (2020).
\bibitem{YLi1}J. Liu, L. Yuan, X. Li, B. Li, K. Zhao, H. Liao, and Y. Li, Gapless spin liquid behavior in a kagome Heisenberg antiferromagnet with randomly distributed hexagons of alternate bonds, Phys. Rev. B \textbf{105}, 024418 (2022).
\bibitem{SLi2}Z. Zeng, X. Ma, S. Wu, H.-F. Li, Z. Tao, X. Lu, X.-h. Chen, J.-X. Mi, S.-J. Song, G.-H. Cao, G. Che, K. Li, G. Li, H. Luo, Z. Y. Meng, and S. Li, Possible Dirac quantum spin liquid in the kagome quantum antiferromagnet YCu$_{3}$(OH)$_{6}$Br$_{2}$[Br$_{x}$(OH)$_{1-x}$], Phys. Rev. B \textbf{105}, L121109 (2022).
\bibitem{YLi2}F. Lu, L. Yuan, J. Zhang, B. Li, Y. Luo, and Y. Li, The observation of quantum fluctuations in a kagome Heisenberg antiferromagnet, Commun. Phys. \textbf{5}, 272 (2022).
\bibitem{Mendels}P. Mendels, F. Bert, M. A. de Vries, A. Olariu, A. Harrison, F. Duc, J. C. Trombe, J. S. Lord, A. Amato, and C. Baines, Quantum magnetism in the paratacamite family: towards an ideal kagome lattice, Phys. Rev. Lett. \textbf{98} 077204(2007).
\bibitem{Helton}J. S. Helton, K. Matan, M. P. Shores, E. A. Nytko, B. M. Bartlett, Y. Yoshida, Y. Takano, A. Suslov, Y. Qiu, J. H. Chung, D. G. Nocera, and Y. S. Lee,  Spin dynamics of the spin-1/2 kagome lattice antiferromagnet ZnCu$_{3}$(OH)$_{6}$Cl$_{2}$, Phys. Rev. Lett. \textbf{98} 107204(2007).
\bibitem{Han}T. H. Han, J. S. Helton, S. Chu, D. G. Nocera, J. A. Rodriguez Rivera, C. Broholm, and Y. S. Lee, Fractionalized excitations in the spin liquid state of a kagome lattice antiferromagnet, Nature \textbf{492} 406(2012).
\bibitem{Shi}Z. L. Feng, Z. Li, X. Meng, W. Yi, Y. Wei, J. Zhang, Y. C. Wang, W. Jiang, Z. Liu, S. Y. Li, F. Liu, J. L. Luo, S. L. Li, G. Q. Zheng, Z. Y. Meng, J. W. Mei and Y. G. Shi, Gapped spin-1/2 spinon excitations in a new kagome quantum spin liquid compound Cu$_{3}$Zn(OH)$_{6}$FBr, Chin. Phys. Lett. \textbf{34} 077502(2017).


\bibitem{Choi}S. Jeon, D. Wulferding, Y. Choi, S. Lee, K. Nam, K. Kim, M. Lee, T.-H. Jang, J.-H. Park, S. Lee, S. Choi, C. Lee, H. Nojiri, and K. Choi, One-ninth magnetization plateau stabilized by spin entanglement in a kagome antiferromagnet, Nat. Phys. \textbf{20}, 1(2024).
\bibitem{Matsuda1}S. Suetsugu, T. Asaba, Y. Kasahara, Y. Kohsaka, K. Totsuka, Boqiang Li, Yuqiang Zhao, Yuesheng Li, M. Tokunaga, and Y. Matsuda, Emergent spin-gapped magnetization plateaus in a spin-1/2 perfect kagome antiferromagnet. Phys. Rev. Lett. \textbf{132}, 226701(2024).     
\bibitem{PALee1}G. Zheng, Y. Zhu, K.-W.Chen, B. Kang, D. Zhang, K. Jenkins, A. Chan, Z. Zeng, A. Xu, O.A. Valenzuela, J. Blawat, J. Singleton, P. A. Lee, S. Li, and L. Li, Unconventional magnetic oscillations in a kagome Mott insulator, PNAS, \textbf{122}, e2421390122(2025).
\bibitem{Matsuda2}S. Suetsugu, T. Asaba, S. Ikemori, Y. Sekino, Y. Kasahara, K. Totsuka, B. Li, Y. Zhao, Y. Li, Y. Kohama, and Y. Matsuda, Gapless spin excitations in a quantum spin liquid state of s=1/2 perfect kagome antiferromagnet (2024), arXiv:2407.16208.
\bibitem{PALee2}G. Zheng, D. Zhang, Y. Zhu, K.-W. Chen, A. Chan, K. Jenkins, B. Kang, Z. Zeng, A. Xu, D. Ratkovski, J. Blawat, A. Bangura, J. Singleton, P. A. Lee, S. Li, and L. Li, Thermodynamic evidence of fermionic behavior in the vicinity of one-ninth plateau in a kagome antiferromagnet (2024), arXiv:2409.05600


\bibitem{Yang1}Jian-Hua Yang and Tao Li, Instability of the U (1) spin liquid with a large spinon Fermi surface in the Heisenberg-ring exchange model on the triangular lattice, Phys. Rev. B \textbf{108}, 235105(2023).
\bibitem{Sorella}S.Yunoki and S. Sorella, Two spin liquid phases in the spatially anisotropic triangular Heisenberg model, Phys. Rev. B \textbf{74}, 014408 (2006).
\bibitem{Nocedal}J. Nocedal and S. J. Wright, $Numerical \ Optimization$, Springer(2008).
\bibitem{Yang2}Jian-Hua Yang and Tao Li, Strong relevance of zinc impurities in spin-1/2 kagome quantum antiferromagnets: A variational study, Phys. Rev. B \textbf{109}, 115103(2024).









 






 

\end{thebibliography}
\end{document}